\documentclass[prb,twocolumn,showpacs,floatfix]{revtex4}    
\usepackage{graphicx}    
\usepackage{amsmath} 
 
\begin{document}    
    
\title{Spin-orbit-enhanced Wigner localization in quantum dots}    
\author{A. Cavalli$^1$, F. Malet$^2$, J. C. Cremon$^2$, and S. M. Reimann$^2$}    
\affiliation{$^1$DTU Nanotech, DTU, DK-2800 Kgs. Lyngby, Denmark   
\\$^2$Mathematical Physics, Lund University, LTH, P.O. Box 118, SE-22100 Lund, Sweden}    
    
\date{\today}    
    
\begin{abstract}    
We investigate quantum dots with Rashba spin-orbit coupling in the    
strongly-correlated regime. We show that the presence of the Rashba     
interaction enhances the Wigner localization in these systems, making     
it achievable for higher densities than those at which it     
is observed in Rashba-free quantum dots. Recurring shapes in the     
pair distribution functions of the yrast spectrum, which might be     
associated with rotational and vibrational modes, are also reported.    
\end{abstract}    
    
    
\maketitle    

\section{Introduction}
    
Electron localization in finite-size nanostructures, reminiscent of the  
Wigner crystallization of the bulk electron gas \cite{Wig34},  
occurs when the density of the system becomes low enough so that the Coulomb  
repulsion dominates over the motion associated with the kinetic energy of 
the electrons. As a consequence, the electrons ``crystallize'', acting as     
classical charges seeking for the equilibrium positions that minimize their     
total energy. From the theoretical point of view, this phenomenon has been  
extensively investigated in e.g. quantum 
dots (QDs) \cite{Mak96,Rei00,Egg99,Fil01} (see also the review Ref. 6),  
quantum wires \cite{Tan98}, and quantum rings \cite{Emp02}, and analogue 
Wigner-localized states have also been observed in other quantum 
systems such as e.g. ultracold atomic gases, ultracold trapped ions, 
complex plasmas, or cuprate chain compounds \cite{Yan07}.  
On the experimental side, signatures of electron localization have been     
found in e.g. quantum dots \cite{Kal08}, quantum wires \cite{Aus05}, or 
carbon nanotubes \cite{Des08}.     
The study of this phenomenon is not only interesting from the purely     
fundamental point of view, but it has also been shown that Wigner-localized     
systems can be useful to design e.g. chains of spatially separated     
quantum bits \cite{Des08} or quantum hard drives \cite{Tay08}.    
    
In experiments with semiconductor quantum dots and wires, however,  
difficulties are encountered when lowering the density down to the  
regimes at which Wigner localization is achieved due to disorder  
effects \cite{Aus05,Kal08}.    
The search for more clear and direct experimental evidence of Wigner    
localization in such systems, as well as for ways that allow its  
observation at more easily accessible densities, is therefore an  
interesting matter of research.     
    
In the context of semiconductor nanostructures, the study of the effects  
of the so-called Dresselhaus \cite{Dre55} (DSOI) and Rashba \cite{Byc84}  
spin-orbit interactions (RSOI) has also attracted great interest in the  
last years. The origin of these interactions is due to the existence of  
internal asymmetries that give rise to net electric fields within the 
heterostructures. In the reference frame of the electrons, 
these electric fields transform into effective magnetic 
fields, which couple to the electronic spins.  
In the case of the Rashba spin-orbit coupling, which is due to the 
asymmetry of the potential well confining the conduction electrons
within the heterostructure, special attention has been paid since  
it was shown that its strength can be experimentally tuned within large  
ranges of values by means of the application of external electric  
fields \cite{Nit97} or by changing the electron density \cite{Mat00},  
making of this mechanism a very promising tool for potential applications  
in e.g. spintronics or quantum computation since it provides a natural  
way to control and manipulate the electronic spin \cite{Wol01}.    
    
Important effects due to the spin-orbit (SO) interaction in semiconductor
nanostructures have been reported on, e.g., the addition energies, the 
electron g-factor, the magnetoconductivity, the spin textures, 
or the spin-relaxation properties \cite{Gov02}.   
Also, spin-orbit-induced electron localization has been observed in 
Coulomb-free closed loops of quantum wires with Rashba SO coupling, 
as well as in quantum rings with both RSOI and DSOI \cite{Ber04}. 
It must be pointed out, however, that this localization is not of 
the Wigner-type since the electrons     
are considered to be non-interacting, but is entirely due to the analogies 
and the interplay of the spin-orbit coupling with external magnetic fields,     
which gives rise to Aharonov-Bohm cages and periodic trapping potentials     
that cause the localization of the particles \cite{Ber04}.    
As a matter of fact, the electron-electron interaction has been taken into  
account in a rather small fraction of the many studies on the spin-orbit  
coupling in semiconductor nanostructures, and only a few have treated it  
without approximations or have considered its interplay with the SO 
terms \cite{Cha05}.  
  
In this Letter, we investigate strongly-correlated InAs quantum dots with 
RSOI and we show that Wigner localization is largely enhanced by the 
presence of the Rashba coupling, making this phenomenon potentially observable 
at sensibly higher electronic densities than those at which it is achieved  
in spin-orbit-free QDs. Since the Dresselhaus contribution is expected to 
be small in narrow-bandgap materials such as InAs \cite{Lom88}, we neglect 
it in the present study.  
    
\section{Model and method}

We consider a two-dimensional InAs quantum dot with parabolic confinement  
in the $xy-$plane given by $V({\bf r})=(1/2)\omega^2(x^2+y^2)$. The Rashba     
spin-orbit coupling is described by the usual term \cite{Byc84}     
\begin{equation}    
H_R=\frac{\hbar k_R}{m^*m_e}\left[p_{y}\sigma_{x}-p_{x}\sigma_{y}\right] \; ,    
\end{equation}    
where $m^*m_e$ is the electron effective mass, with $m^*=0.023$, $p_i$ is  
its linear momentum in the $i-$direction, $\sigma_i$ is the corresponding  
Pauli matrix, and $k_R$ determines the strength of the interaction. 
%In real  
%experiments, values for $\hbar^2 k_R/(m^*m_e)$ up 
%to $\sim 3\times 10^{-11}$ eV$\cdot$m have been achieved \cite{Sat01,Mat00}.    
    
The QD is considered to have $N$ electrons interacting with each other  
through the Coulomb interaction. Expressing the energies and lengths  
in units of $\hbar\omega$ and $l_{\omega}=\sqrt{\frac{\hbar}{m^*m_e\omega}}$  
respectively, the full Hamiltonian of the system, in second quantization  
form, reads    
\begin{eqnarray}    
H & = & \sum_{\alpha\beta}\Big\langle    
\alpha\mid\frac{p_{x}^{2}}{2}+\frac{p_{y}^{2}}{2}    
+\frac{1}{2}\left(x^{2}+y^{2}\right)    
\nonumber \\    
& + & \widetilde{k_R}\left[p_{y}\sigma_{x}-p_{x}\sigma_{y}\right]\mid    
\beta\Big\rangle \,a_{\alpha}^{\dagger}a_{\beta}    
\nonumber \\    
 & + & \frac{1}{2}\sum_{\alpha\beta\gamma\delta}\langle \alpha,\,    
 \beta\mid\frac{g}{\mid\mathbf{r}-    
\mathbf{r}'\mid}\mid \gamma,\, \delta\rangle    
\,a_{\alpha}^{\dagger}a_{\beta}^{\dagger}a_{\gamma}a_{\delta}\; ,    
\end{eqnarray}    
where $a_{\mu}^{\dagger}$, $a_{\mu}$ ($\mu=\alpha,\beta,\gamma,\delta$)  
are the usual creation and annihilation operators for the single-particle  
state $|\mu\rangle$, and $\widetilde{k_R}\equiv l_{\omega}k_R$ is a  
dimensionless parameter determining the effective spin-orbit interaction  
strength. The prefactor in the Coulomb term is given by    
$g=e^2/(4\pi\epsilon_0\epsilon_r l_{\omega}\hbar\omega) = l_{\omega}/a_B^*$,
where $a_B^*$ is the effective Bohr radius, which for InAs is approximately  
34 nm. The confinement frequency is related to the average electron  
density $n$ through the relation  
$\omega^2=\hbar^2/(m^2a_B^*\sqrt{N}r_s^3)$, \cite{Rei00} where   
$r_s=1/\sqrt{\pi n}$ is the Wigner-Seitz radius. 
    
Using the configuration-interaction method, we numerically find 
the many-particle eigenstates of the system, expanded in a basis 
of Slater determinants made up of 2D harmonic-oscillator  
single-particle orbitals $|\mu\rangle=|n,m,s\rangle$,  
where $n$, $m$ and $s=\uparrow,\downarrow$ are, respectively,  
the radial, azimuthal, and spin quantum numbers. 
Defining the equivalent set $n_{\pm}\equiv (2n+|m|\pm m)/2$,  
one has for the matrix elements of the Rashba term:  
\begin{eqnarray} 
&\langle n^{\prime},\, m^{\prime},\uparrow\mid H_R 
\mid n,\, m,\uparrow\rangle 
=\langle n^{\prime},\, 
m^{\prime},\downarrow\mid H_R 
\mid n,\, m,\downarrow\rangle=0, 
\nonumber\\ 
&\langle {n_{+}}^{\prime},{n_{-}}^{\prime},\uparrow\mid H_R\mid n_{+}, 
n_{-},\downarrow\rangle =  -\widetilde{k_R}\times 
\nonumber\\ 
&\left[\sqrt{n_{-}+1}\delta_{n_{+},{n_{+}}^{\prime}} 
\delta_{(n_{-}+1),{n_{-}}^{\prime}}-\sqrt{n_{+}}\delta_{(n_{+}-1),{n_{+}}^{\prime}} 
\delta_{n_{-},{n_{-}}^{\prime}}\right], 
\nonumber\\ 
&\langle {n_{+}}^{\prime},{n_{-}}^{\prime},\downarrow\mid H_R\mid n_{+},  
n_{-},\uparrow\rangle =  \widetilde{k_R}\times 
\nonumber\\ 
&\left[\sqrt{n_{+}+1}\delta_{(n_{+}+1),{n_{+}}^{\prime}} 
\delta_{n_{-},{n_{-}}^{\prime}}-\sqrt{n_{-}}\delta_{n_{+},{n_{+}}^{\prime}} 
\delta_{(n_{-}-1),{n_{-}}^{\prime}}\right] . 
\end{eqnarray} 
The (in principle) infinitely large basis space is truncated, 
taken to be large enough to ensure converged solutions, which 
in our calculations are achieved with ~15 harmonic oscillator 
shells. As the total angular momentum $J_z=L_z+S_z$ represents 
a good quantum number, the diagonalization is in practice done 
separately for each considered $J_z$. 
 
A standard value for the confinement energy is of the order  
of $\hbar\omega\sim 5$ meV, with associated effective Coulomb  
strength $g=0.75\equiv g_0$. We have studied the interval  
$g_0\le g\le 3 g_0$, corresponding to  
5 meV $\le \hbar\omega\le 0.56$ meV. 
Finally, regarding the RSOI term, we have considered the range 
$0.263\le \widetilde{k_R}\le 2.63$, and we have found  that the 
spin-orbit-induced localization effects become clearly observable 
for strengths of the Rashba coupling constant 
$\hbar^2k_R/m\gtrsim 8.5\times 10^{-11}$ eV$\cdot$m.  
While being slightly more than two times larger than the ones 
experimentally accessible nowadays \cite{Mat00,Sat01}, such values 
should be reachable in the near future considering the ongoing 
research on new materials with larger SO couplings \cite{Hao10}. 
    
\section{Results}

When the RSOI is neglected ($k_R=0$), we find that the electrons 
start to localize only for strengths of the Coulomb interaction 
larger than $\sim 5 g_0$, corresponding to $r_s\gtrsim 4.4 a_B^*$, 
as expected \cite{Rei00,Egg99}.  
For weaker interactions, we only observe a slight deformation of  
the electron cloud as a whole. Also, when the localization occurs,  
there is a depletion of the electron density in the center  
of the trap due to the repulsion between the particles.  
 
This scenario, however, sensibly changes when the Rashba term is taken  
into account. In Figs. 1 and 2 we show the pair distribution function     
\begin{equation}    
PDF({\bf r},{\bf r}')= 
\langle\hat{\psi}^{\dagger}({\bf r})\hat{\psi}^{\dagger}({\bf r}') 
\hat{\psi}({\bf r}')\hat{\psi}({\bf r})\rangle\; ,    
\end{equation}    
which gives the probability of finding a particle at the position  
${\bf r}$ provided that another one is at ${\bf r}'$, for the 3- and  
4-electron quantum dots for different strengths of the effective  
spin-orbit and Coulomb interactions. 
One can clearly see that the inclusion of the Rashba term, and 
especially for large values of $\widetilde{k_R}$, significantly 
enhances the electron localization. Indeed, in both cases it is 
already visible for $\widetilde{k_R}=1.84$ and $g=3 g_0$, 
corresponding to $\hbar\omega=$ 0.56 meV 
and $\hbar^2k_R/m=8.5\times 10^{-11}$ eV$\cdot$m, with the 
electrons distributing themselves on the vertices of an  
equilateral triangle and of a square, respectively, 
analogously to what is observed in quantum dots submitted 
to perpendicular magnetic fields \cite{Mak96,Kal08}. This 
corresponds to Wigner-Seitz radii of $r_s=2.5 a_B^*$ for 
$N=3$ and to $r_s=2.4 a_B^*$ for $N=4$, i.e. almost half 
the values at which localization is observed in spin-orbit-free 
quantum dots. A further increase of the RSOI while keeping 
the confinement strength fixed leads to a more pronounced 
localization of the electrons, as can be seen from the figures 
when $\widetilde{k_R}=2.63$. 
 
We have also investigated a dot with six electrons, although in this 
case an optimal convergence of the numerical simulations has not been 
possible to achieve. Nevertheless, we have observed that the behavior 
of the system is qualitatively similar to the cases with $N=3$ and 
$N=4$, with localized configurations appearing at similar effective 
Coulomb strengths with $r_s\sim 2.2 a_B^*$, made up of five electrons 
forming a pentagon with the sixth one placed at the center of the dot. 
Finally, we also want to mention that in some papers electron 
localization has been addressed making use of quantitative suitable 
criteria \cite{Fil01,Bon08}. In our study, however, we have limited 
ourselves to the rather qualitative discussion presented above.

\begin{figure}[h!]    
\includegraphics[width=1\columnwidth]{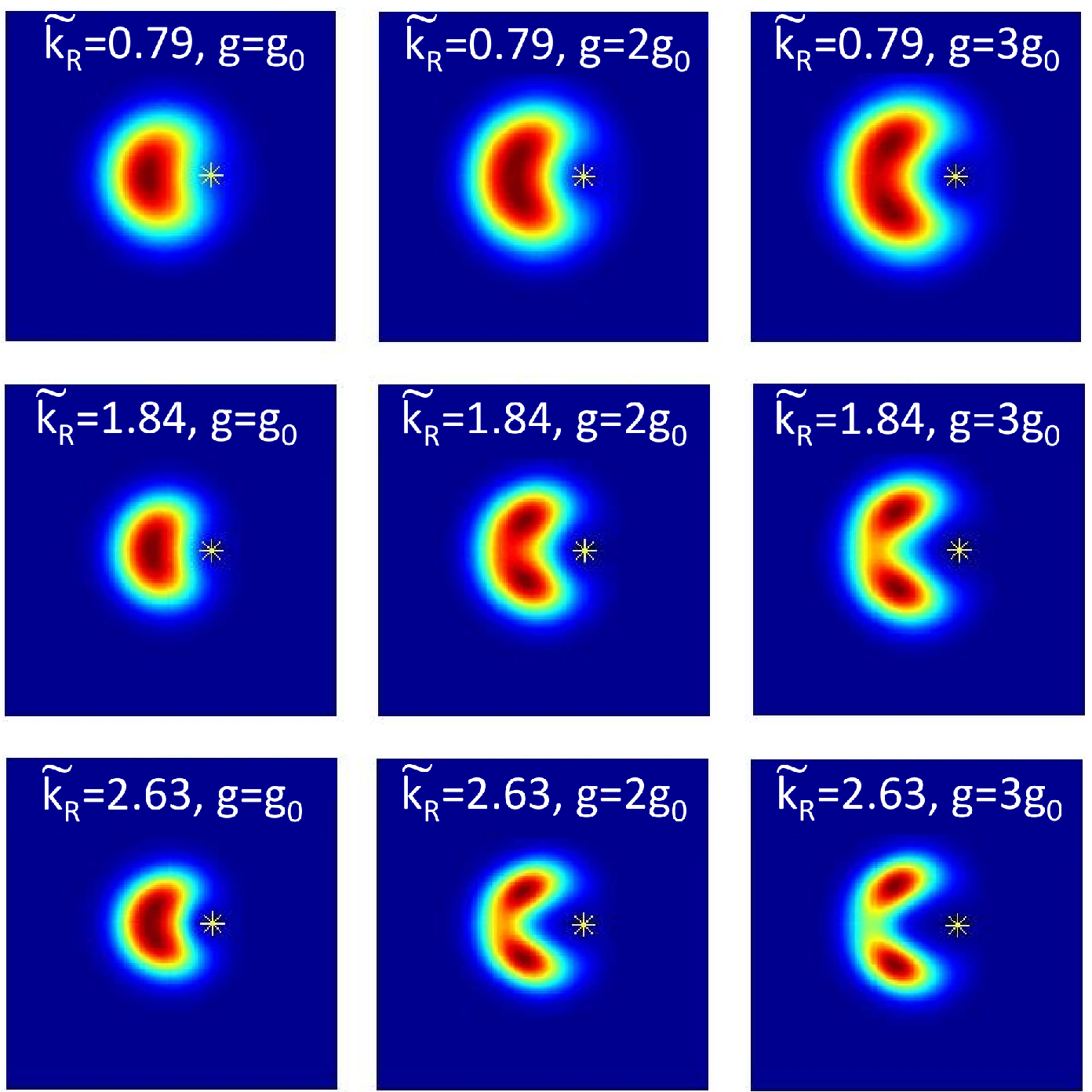}    
\caption{Pair distribution functions for the 3-electron quantum dot and for     
different strengths of the effective Rashba (increasing from top to bottom)  
and Coulomb (increasing from left to right) interactions. The star symbol     
indicates the position of the reference electron.}    
\label{fig1}    
\end{figure}    
    
\begin{figure}[h!]    
\includegraphics[width=1\columnwidth]{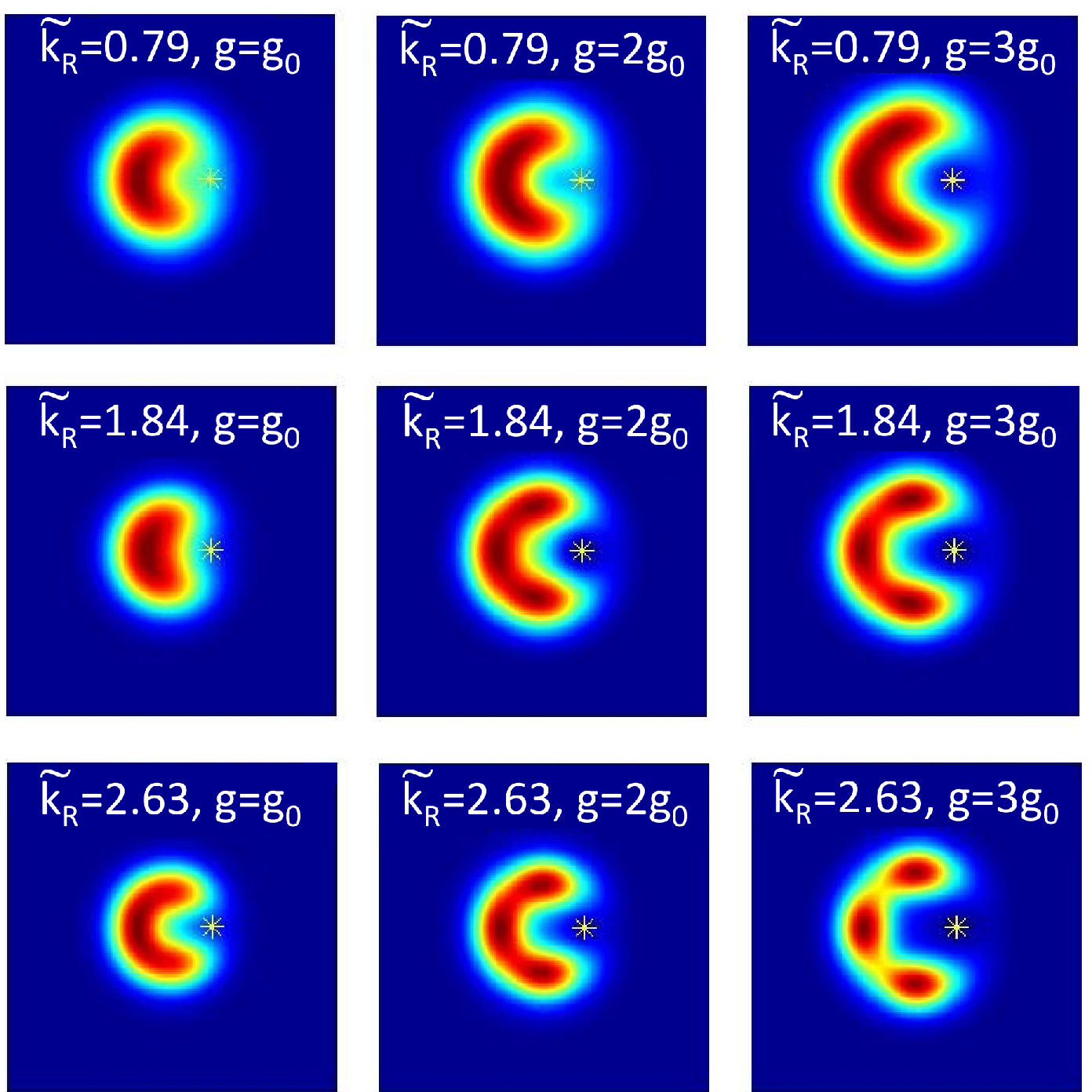}    
\caption{Same as Fig. 1 for the quantum dot with $N=4$.}    
\label{fig2}    
\end{figure}    
    
The enhancement of the Wigner localization due to the Rashba 
spin-orbit coupling can be qualitatively explained already 
from the non-interacting-electron picture. Indeed, we have 
observed that the spatial extension of the single-particle 
density profiles is reduced when the RSOI term is taken into 
account as compared to the harmonic oscillator case with 
$k_R=0$. This is in close analogy with the Fock-Darwin 
single-particle states of quantum dots under magnetic 
fields \cite{Mak90} and may be understood from the nature 
of the spin-orbit interaction described above. Therefore, 
it can be concluded that one of the effects of the Rashba 
coupling is to minimize the overlapping between the different 
electrons and thus to favour their spatial separation with 
respect to each other.     
    
It is also interesting to study the so-called yrast 
spectrum \cite{yrastnote}, i.e., the lowest energy level 
for a given total angular momentum, which we plot in Fig. 3 for the 
three-electron dot in different parameter regimes. This again shows very 
close analogy with what is observed in quantum dots under perpendicular 
magnetic fields \cite{Mak96}, the main difference being in the 
low $J_z$ range. In the Rashba case, low $|j_z|$ single-particle states 
are almost degenerate and therefore the spectrum is rather flat for 
low values of $J_z$. On the contrary, under an applied magnetic field, 
a preferred spin direction is selected and low $J_z$ states, realized 
using s.p. states of opposite angular momentum, are energetically 
unfavourable. This explains the overall parabolic shape of the yrast 
profiles shown in \cite{Mak96}, different from the almost degenerate 
low $J_z$ levels in the Rashba case. For higher angular momenta, 
however, the magnetic and Rashba yrast spectra become more similar: 
in both cases the yrast lines show oscillations with period $N$, 
which become more pronounced as the Coulomb repulsion increases. 
The presence of these oscillations is explained in Ref. 2 by the 
vanishing of the exchange energy term for configuration with electrons 
occupying adjacent $j_z$ single-particle levels, which occur only for 
every third unit of angular momentum. 
Moreover, for the strong magnetic field case --and thus for high $J_z$-- 
it can be seen that the electrons behave as a rigid, molecule-like
configuration \cite{Mak96,Nik07}. A semiclassical interpretation 
of the yrast spectrum is then possible, with the local minima 
corresponding to purely rotational modes and with the intermediate 
states containing at least one vibrational quantum \cite{Kos01}.
  
\begin{figure}[t]    
\includegraphics[width=1\columnwidth]{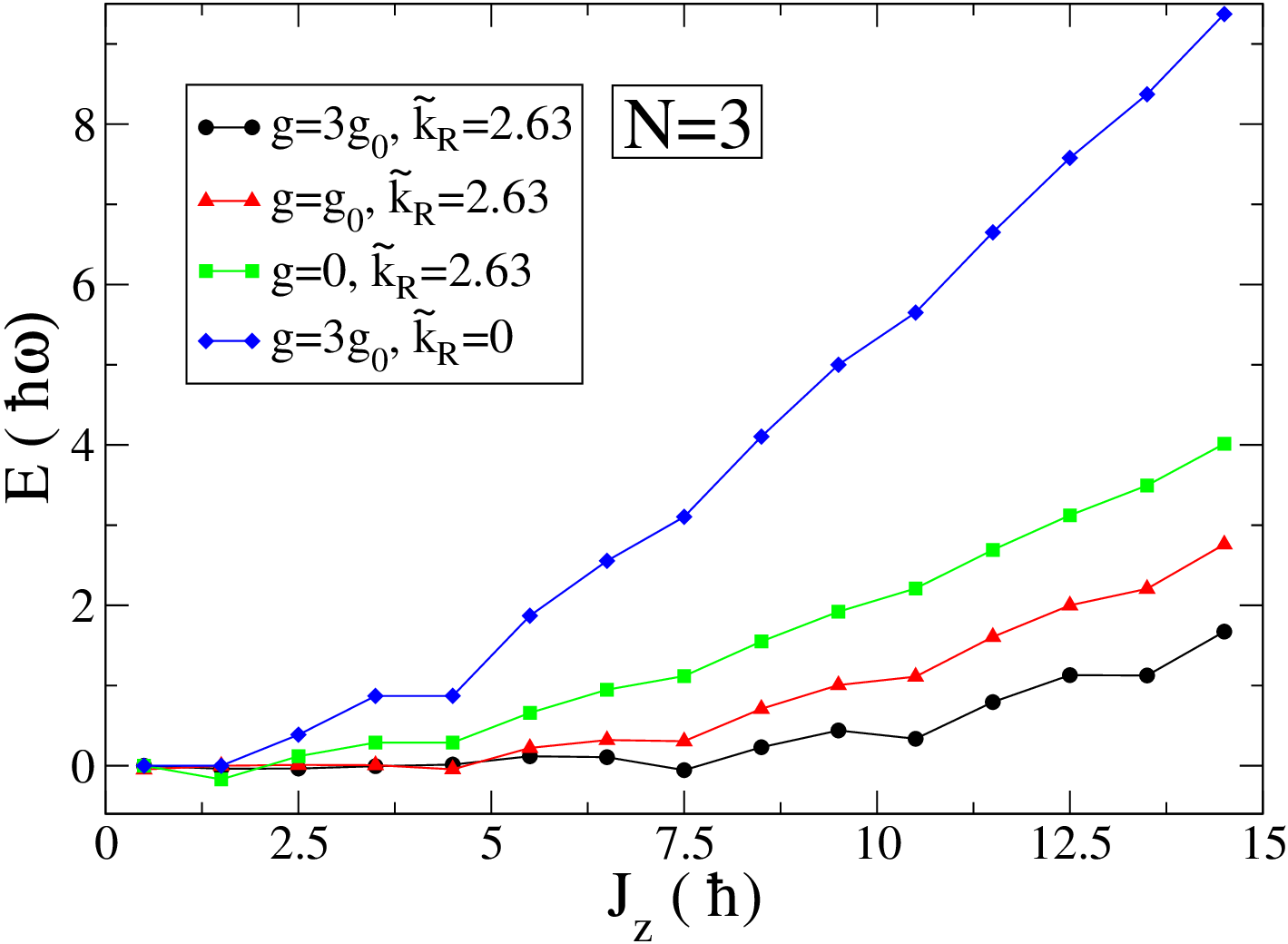}    
\caption{Yrast spectrum for the 3-electron quantum 
dot with $g=3 g_0$ and $\widetilde{k_R}=0$ (blue 
diamonds), $g=0$ and $\widetilde{k_R}=2.63$ (green 
squares), $g=g_0$ and $\widetilde{k_R}=2.63$ (red 
triangles), and $g=3 g_0$ and $\widetilde{k_R}=2.63$ 
(black circles). The energies have been shifted to
have the same value at $J_z=0.5 \hbar$ for a better 
comparison.}
\label{fig3}    
\end{figure}    
    
Finally, a qualitative signature of the above-mentioned molecular
behaviour in the Rashba-interacting case can be observed from Fig. 4, 
where we show the pair distribution functions corresponding 
to the Wigner-localized case for different values of the total angular 
momentum. One can recognize recurring shapes ($J_z=7.5 \hbar$ 
and $J_z=10.5 \hbar$, $J_z=8.5 \hbar$ and $J_z=11.5 \hbar$, $J_z=9.5 \hbar$ 
and $J_z=$12.5), which might be associated with specific rotational 
and vibrational states for localized electrons in quantum dots. 
Indeed, the standard triangular shapes are found at the local minima, 
consistently with the hypothesis that they correspond to rotational modes, 
where no deformation is involved. Other shapes also appear periodically, and 
they look consistent with the vibrations of a triangular molecule (stretching 
and shrinking the triangle basis, for example, similarly as patterns 
previously observed \cite{Nik07}). We think that the appearance of these 
deformed pair distribution functions matching the periodicity of the yrast 
spectrum is a strong hint of a ``rigid-molecule'' behavior of the electrons. 

\begin{figure}[t]    
\includegraphics[width=1\columnwidth]{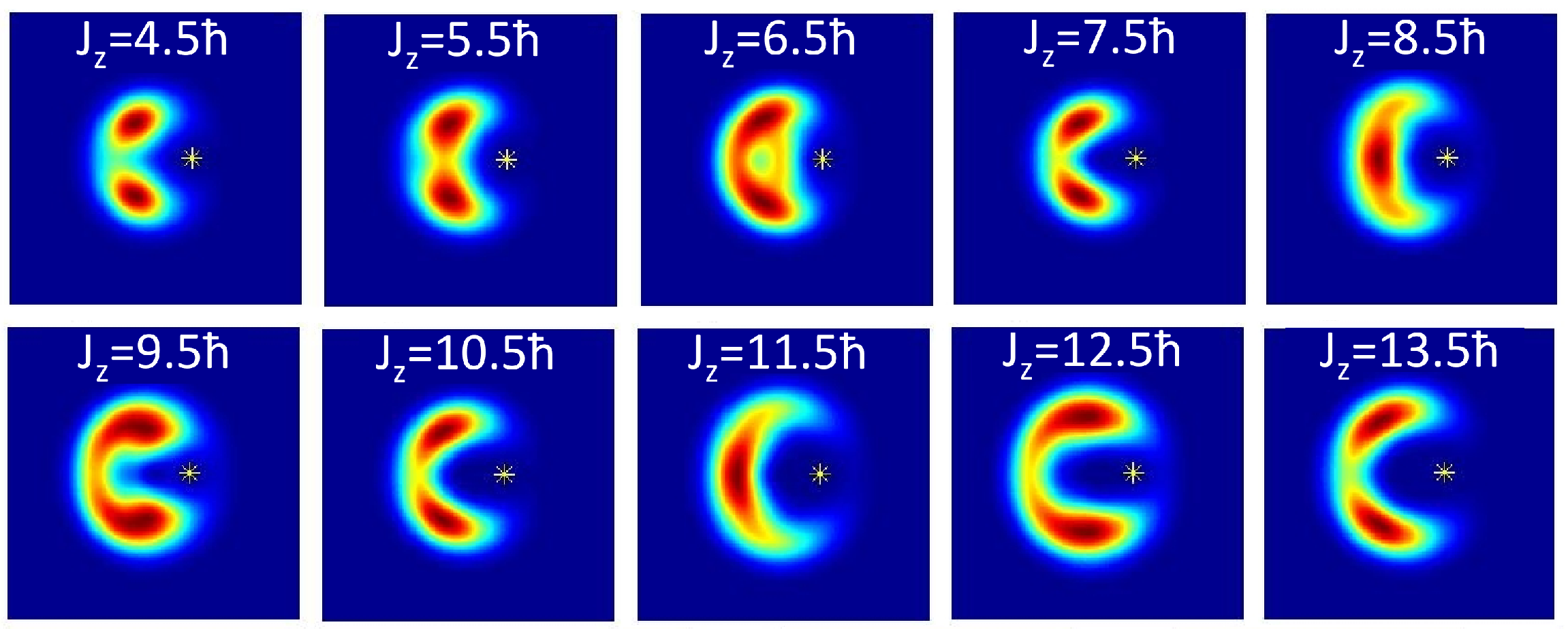}    
\caption{Pair distribution functions for different values 
of $J_z$ for the three-electron dot in the Wigner-localized 
regime ($g=3 g_0$ and $\widetilde{k_R}=2.63$).}     
\label{fig4}    
\end{figure}    
    
\section{Conclusions}

In conclusion, we have found that the presence of Rashba spin-orbit  
coupling in strongly-correlated quantum dots largely enhances the 
Wigner localization in these systems and makes it observable at 
higher densities than in the spin-orbit-free case. We report several 
analogies with quantum dots under magnetic fields, and in particular 
we show that the pair distribution functions corresponding to the 
states of the yrast spectrum of the Wigner-localized dot present 
recurring shapes that one can associate with rotational and 
vibrational modes. Our results thus point out the importance of 
the experimental investigation of Wigner localization in materials 
with large spin-orbit coupling. Although not modeled here, any 
disorder in the material can be expected to contribute to further 
enhance the localization \cite{Aus05,Kal08}. The effects of the 
Dresselhaus spin-orbit interaction on the electron localization, 
and in particular its interplay with the Rashba term, is also an 
interesting subject that will be investigated in future works.    

\section{Acknowledgments}
    
We thank M. Koskinen, H. Linke, M. Manninen, and H.Q. Xu for 
discussions and valuable comments. This work was financed by 
the Swedish Research Council. We also thank the nmC@LU for
support.

\end{document}